\documentclass[aps,prl,superscriptaddress,reprint]{revtex4-2}

\usepackage[utf8]{inputenc}
\usepackage{geometry}
\usepackage{xcolor}
\usepackage{pst-node}
\usepackage{graphicx}
\usepackage{amsmath}
\usepackage{mathtools}
\usepackage{kbordermatrix}
\usepackage{hyperref}
\usepackage[shortlabels]{enumitem}
\usepackage{dsfont}
\usepackage{amsthm}
\usepackage{amssymb}
\usepackage{amstext}
\usepackage{amsfonts}
\usepackage{nicefrac}
\usepackage{extarrows} 
\usepackage{graphicx}
\usepackage{relsize}

\renewcommand{\kbldelim}{(}
\renewcommand{\kbrdelim}{)}

\renewcommand{\phi}{\varphi}
\renewcommand{\epsilon}{\varepsilon}

\DeclareMathOperator{\Id}{Id}

\newcommand{\R}{\mathbb{R}}
\newcommand{\C}{\mathbb{C}}

\renewcommand{\H}{\mathcal{H}}
\newcommand{\F}{\mathcal{F}}
\newcommand{\norm}[1]{\left\lVert#1\right\rVert}

\newtheorem{theorem}{Theorem}

\newtheorem*{theorem*}{Theorem}

\begin{document}

\title{Exact matrix product state  representation and convergence \\ of a fully correlated electronic wavefunction
in the infinite basis limit}
\author{Gero Friesecke}%
 \email{gf@ma.tum.de}
\affiliation{%
  Department of Mathematics, Technical University of Munich, Germany
}%

\author{Benedikt R. Graswald}
\email{benedikt.graswald@ma.tum.de }
\affiliation{%
  Department of Mathematics, Technical University of Munich, Germany
}%

\author{Örs Legeza}
\email{legeza.ors@wigner.hu }
\affiliation{%
Wigner Research Centre for Physics, H-1525, Budapest, Hungary
}%
\affiliation{
Institute for Advanced Study,Technical University of Munich, Germany
}

\date{\today}

\begin{abstract}
\noindent \textbf{Abstract.}
In this article
we present the exact representation of a fully correlated electronic wavefunction as the single-particle basis approaches completeness. It consists of 
a half-infinite chain of  
matrices of exponentially increasing size. 
The complete basis limit is illustrated numerically using the density matrix renormalization group method by computing the core-valence entanglement in the 
C$_2$ ground state in increasing subsets of cc-pVTZ and pVQZ bases until convergence is reached. 
\end{abstract}

\maketitle

 \section{Introduction}
To achieve controllable quantum devices, 
 understanding the intrinsic quantum entanglement hidden within the fundamental molecular building blocks of nature has been a central goal in the past decades \cite{hettich2002nanometer, lin2020quantum}.
 Fully resolving this entanglement in theory requires
 not just no limitations on excitation ranks, but also -- owing to the 
 fact that a single electron can assume  infinitely many quantum states -- reaching the complete basis set (CBS) limit.
The difficulty of this latter aim in correlated models was emphasized many decades ago by Pople \cite{Pople73}; due to the curse of dimensionality progress has been limited to  a fixed excitation level (and global quantities like energy) \cite{elm2017, HALKIER1998, nagy2019, varandas2021}. 

Recent advances in methods based on matrix product state (MPS)  factorization have made it possible to calculate  electronic states of small molecules accurately in large basis sets compared to conventional methods \cite{chan2011density,szalay2015tensor,baiardi2020density}. 
However, the question of what happens to the MPS representation in the CBS limit has remained open both theoretically and numerically.

Here we prove that electronic wavefunctions 
possess an exact MPS representation in a complete basis. It is given by a half-infinite chain of matrices and is thus loosely reminiscent of certain infinite MPS that have been used in solid state physics \cite{vidal2007, Jordan2008}, but here the size of the matrices is found to increase exponentially along the chain.
We also demonstrate this result numerically by computing the 
dicarbon ground state in increasingly large size-$K$ subsets of cc-pVTZ and pVQZ bases (108 and 216 spin orbitals, respectively) 
and exhibiting the convergence of the MPS matrices as $K$ is increased. 

To the best of our knowledge, our work is the first 
demonstration of convergence of a fully correlated wavefunction with more than two electrons in the CBS limit.
The entanglement for a bipartite decomposition of a pure state is completely characterized by the Schmidt spectrum, i.e., the associated singular values.
When one of the parts is finite-dimensional and the other is infinite-dimensional, the Schmidt spectrum still consists of finitely many values only, and can be numerically computed. Here we do so for the finite-dimensional 4-electron core space of C$_2$ and the  infinite-dimensional valence space, in which case there are exactly $256$ Schmidt values, all of which are accurately calculated and demonstrated to converge.

Our work sheds new light on quantum entanglement in molecules by representing the exact (complete basis) many-electron wavefunction introduced by Dirac \cite{dirac1929quantum} as an infinite product, 
which allows us to overcome the cutoff on the excitation level and finally bring together theory and computational feasibility. By contrast, the fundamental work of L{\"o}wdin \cite{lowdin1955quantum} on Full-CI in the CBS limit, which leads to an infinite series representing the wave function,
remained computationally elusive.

\section{ Set-up and notation} Let $\H$ be an infinite-dimensional separable Hilbert space spanned by orthonormal orbitals 
 $\{\phi_i\}_{i=1}^{\infty}$. Let ${\mathcal V}_N$ be the $N$-fold antisymmetric product $\bigwedge_{i=1}^N \H$, and let $\F$ be the ensuing Fock space, 
 \[
    \F := \bigoplus_{N=0}^\infty {\mathcal V}_N.
 \]
 The prototypical case we have in mind is that $\H$ is the full (untruncated) Hilbert space $\H=L^2(\R^3;\C^2)$ of square-integrable single-electron wavefunctions, in which case ${\mathcal V}_N$ is the space of square-integrable antisymmetric $N$-electron wavefunctions. 
 We denote the Fock space over the  basis-truncated Hilbert space spanned by the first $k$ orbitals by $\F_k$.
 Slater determinants indexed by their binary label are denoted $\Phi_{\mu_1...\mu_K}$, that is to say
\begin{align} \label{eq:sdbin}
    \Phi_{\mu_1...\mu_K} := |\varphi_{i_1}...\varphi_{i_N}\rangle \mbox{ if }\mu_{i}=1 \mbox{ exactly }\nonumber  \\ \mbox{when }i\in\{i_1,...,i_N\},~ i_1<\ldots<i_N. 
\end{align}

\section{ MPS for infinite basis}
\begin{figure}[ht!]
    \centering
    \includegraphics[width = 0.455 \textwidth]{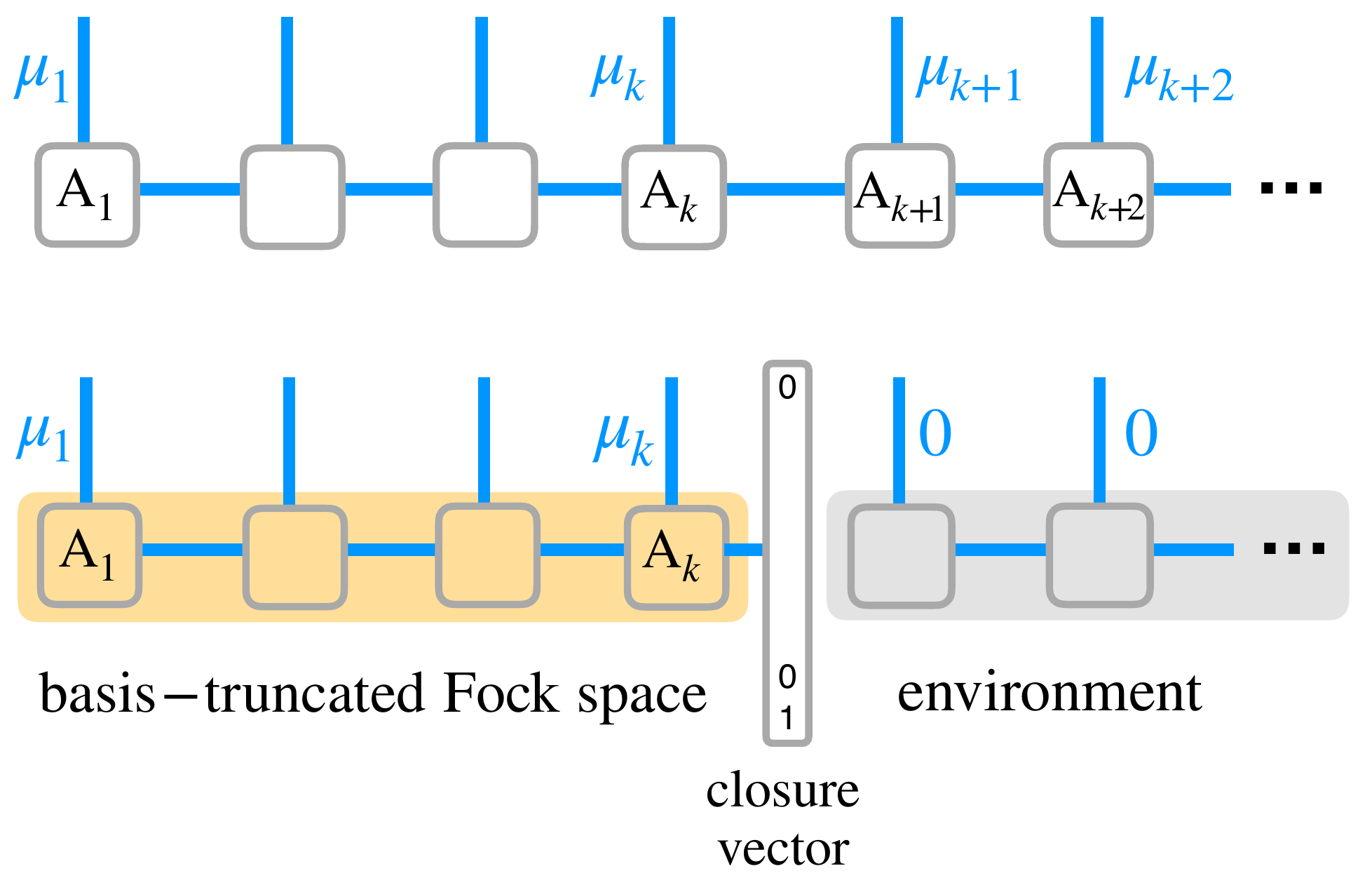}
    \vspace*{-5mm}
    
    \caption{{\it Top:} MPS representation of a quantum state in the full infinite-basis Fock space,   eq.~\eqref{eq:prop_inf_mps}. {\it Bottom:} Inserting a closure vector after the $k^{th}$ node tensor filters out its  projection onto the Fock space over the first $k$ orbitals, eq.~\eqref{eq:proj_inf}, and corresponds to setting the physical variables associated to all higher orbitals to zero.}
    \label{fig:mps_infinite}
\end{figure}

For a finite basis set (or finite spin chain), every quantum state in the Fock space has an exact MPS representation provided the size of successive MPS matrices is allowed to increase exponentially until the middle  basis function (or site) before decreasing again (see e.g. \cite{schollwock2011density}). We now  construct an analogous exact MPS representation for
\textcolor{black}{every} 
quantum state in the Fock space over an  infinite basis set (or half-infinite spin chain). In this case 
one obtains an infinite sequence of succesive MPS matrices of exponentially increasing size; to recover the state, which is a scalar, one has to insert a closure vector after the first $k$ matrices and take the limit $k\to\infty$. See Fig. \ref{fig:mps_infinite}. More precisely:

\begin{theorem}[MPS for infinite basis]\label{thm:mps_infinite}
 Given any normalized quantum state $\Psi$ belonging to the full infinite-basis Fock space $\F$, there exist left-normalized tensors $A_k \in \C^{2^{k-1} \times 2 \times 2^{k}}$ such that 
 \begin{align}
 \label{eq:prop_inf_mps}
 \lim_{k\to\infty}  \sum_{\mathclap{\mu_1,\ldots,\mu_k }}  A_1[\mu_1] ... A_k[\mu_k] \!
   \begin{pmatrix} 0  \\ \vdots \\ 0 \\ 1 \end{pmatrix} \! \Phi_{\mu_1 ... \mu_k} = \Psi.
 \end{align}  
 Here the sums run over $\mu_1,...,\mu_k\in\{0,1\}$, and $\Phi_{\mu_1 ... \mu_k}$ is given by \eqref{eq:sdbin}. Moreover for every $k$,
\begin{align} \label{eq:proj_inf}
 \!\!\!\sum_{\mathclap{\mu_1, ...,\mu_k}}  
    A_1[\mu_1] ...  A_{k}[\mu_k]
    \begin{pmatrix} 0  \\ \vdots \\ 0 \\ 1 \end{pmatrix}
    \!
    \Phi_{\mu_1 \ldots \mu_k} \!
    =
   \frac{ P_{\F_k} \Psi}{ \mathsmaller{\norm{P_{\F_k}\Psi}}  } 
\end{align}
where $P_{\F_k}$ denotes the projector onto the Fock space over the first $k$ basis functions, and we use the convention $\frac{v}{||v||}=0$ when $v=0$.
 \end{theorem}

\begin{figure}[ht!]
    \centering
    \includegraphics[width=0.52\textwidth]{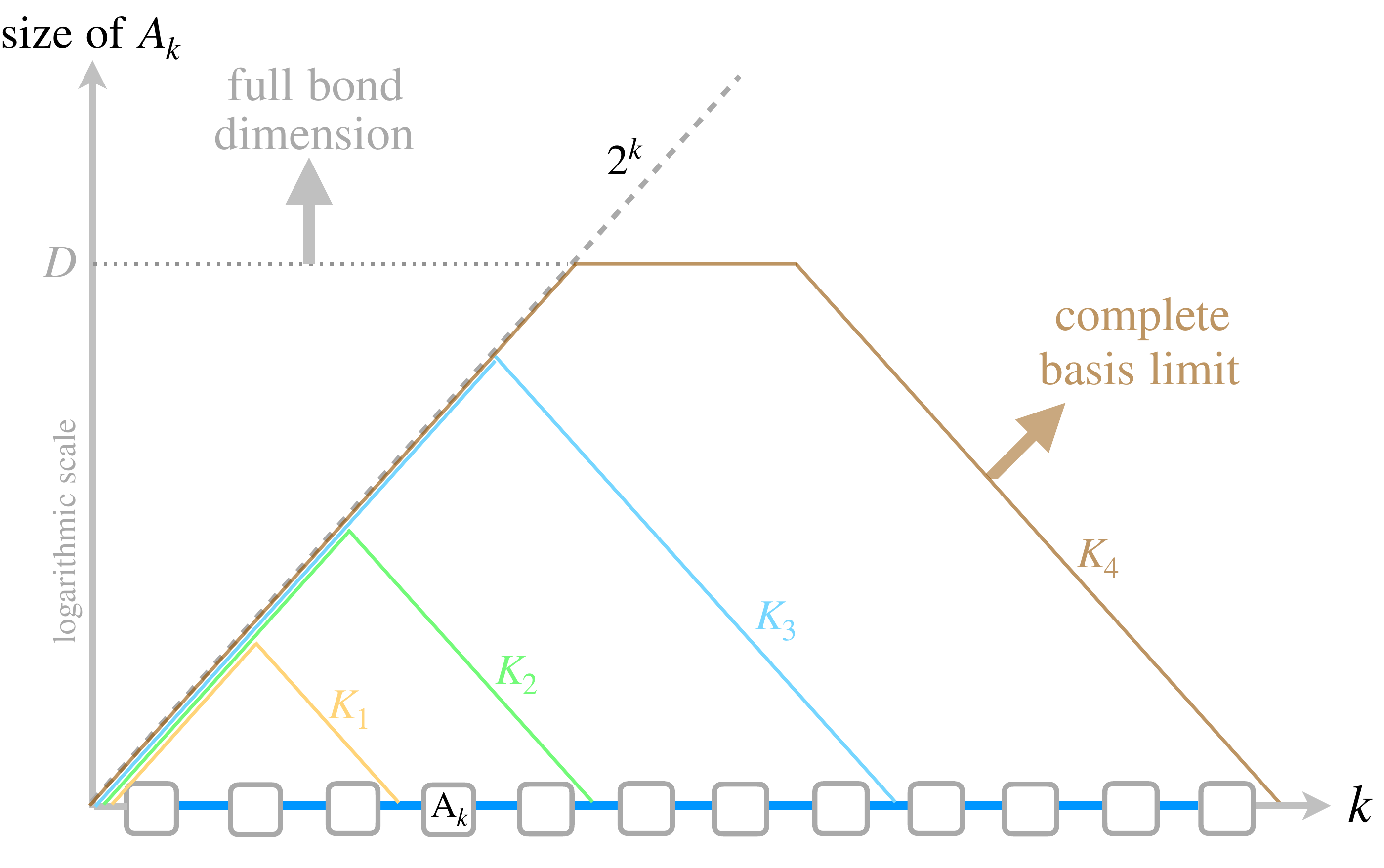}
    \caption{Construction of the exact MPS matrices $A_k$ of a quantum state associated with the 
    basis functions $\varphi_k$ in the complete basis ($K\to\infty$) limit, using basis-truncated approximations with $K$ basis functions. The $A_k$ have size $\min(2^k,2^{K-k})$ in the finite basis (colored lines) and hence maximal size $2^k$ (dashed line); in the numerical results we also used increasing finite-$D$ cutoffs (dotted line) until convergence.}
    \label{fig:matrices}
\end{figure}

 \begin{proof}
The idea is to approximate $\Psi$ by states $\Psi_K$ in the basis-truncated Fock space $\F_K$, obtain corresponding $K$-dependent MPS matrices $A_1^{_{(K)}}\! , \, A_2^{_{(K)}}\! , ...$, use the gauge freedom to achieve eq.~\eqref{eq:proj_inf} for $\Psi_K$ in place of $\Psi$ when $k<K$, and let the basis truncation parameter $K$ go to infinity to arrive at  universal ($K$-independent) matrices $A_1, \, A_2, ...$ (see Fig. \ref{fig:matrices}). 

Now in more detail, start from any sequence $\Psi_K$ of approximations to $\Psi$ belonging to the basis-truncated Fock space $\F_K$ such that $\lim_{K\to\infty}\Psi_K=\Psi$. \textcolor{black}{Such approximations always exist; for instance one could take $\Psi_K$ to be the orthogonal projection  of $\Psi$ onto $\F_K$. Alternatively, }when $\Psi$ is the unique ground state of some Hamiltonian $H$, $\Psi_K$ could be taken to be any ground state of $H$ in $\F_K$, see the next section. It is a
standard fact
(see e.g. \cite{schollwock2011density}) that $\Psi_K$ can be represented in MPS form,
 \begin{align} \label{eq:finite}
 \sum_{\mathclap{\mu_1, ...,\mu_K}}  A^{(K)}_1[\mu_1] \ldots  A^{(K)}_K[\mu_K] \norm{\Psi_K}
 \Phi_{\mu_1 \ldots \mu_K },
 \end{align}
 where the matrices $A^{(K)}_k[\mu_k]$ are left normalized, 
 \begin{equation}\label{eq:LN}
 \sum_{\mu_k=0}^1 
 \bigl(A^{(K)}_k[\mu_k]
 \bigr)^\dagger
 A^{(K)}_k[\mu_k]  
 =
 \Id, 
 \end{equation}
and of size $2^{k-1}\times 2^k$ for $k\le K/2$. 
We use the remaining gauge freedom 
\[
    \big(A_k^{(K)}[\mu], A_{k+1}^{(K)}[\mu']\big) 
    \mapsto  
    \big(A_k^{(K)}[\mu]Q, Q^\dagger A_{k+1}^{(K)}[\mu'])
\]
with $Q$ unitary to achieve that the vector 
$A_{k+1}^{_{(K)}}[0]\ldots A_{K}^{_{(K)}}[0]=:v_k^{_{(K)}}$
has the normal form
\begin{equation}\label{eq:vklnorm}
    v_k^{(K)} = \norm{ v_k^{(K)}}
    \mathbf{e}_k,
   \end{equation}
where $\mathbf{e}_k = (0,\ldots,0,1) \in \R^{2^k}$ is the closure vector from the theorem.
Replacing the matrix product 
$A_{k+1}^{_{(K)}}[\mu_{k+1}]\ldots A_{K}^{_{(K)}}[\mu_K]$ in \eqref{eq:finite} by its value at $\mu_{k+1}=\ldots=\mu_{K}=0$ and using \eqref{eq:vklnorm}, we have for any $k<K$
\begin{align*}
\sum_{\mathclap{\mu_1, ...,\mu_K } }
A^{(K)}_1[\mu_1] \ldots  A^{(K)}_{k}[\mu_k]
\norm{\Psi_K} 
&v_k^{(K)}
\Phi_{\mu_1\ldots \mu_k}\\[-2mm]
&= \, P_{\F_k} \, \Psi_K,
\end{align*}
where $P_{\F_k}$ is the orthogonal projector onto the Fock space over the first $k$ basis functions. In particular, $||{\Psi_K}|| \cdot ||v_k^{_{(K)}}|| = ||P_{\F_k} \Psi_K||$. Combining this with  \eqref{eq:vklnorm} yields
\begin{align}\label{eq:projfinite} \nonumber 
   \norm{P_{\F_k}\Psi_K}  \sum_{\mathclap{\mu_1, ...,\mu_k }}
    A^{(K)}_1[\mu_1] ...  A^{(K)}_{k}&[\mu_k]
    \mathbf{e}_k
    \Phi_{\mu_1 ... \mu_k}\\[-2mm]
    &=
    P_{\F_k} \Psi_K .
\end{align}

The idea now is to fix $k$ and let $K\to \infty$. For $K\geq 2k$ the matrices $A_k^{_{(K)}}$ have fixed size $2^{k-1}\times 2^k$, and are uniformly bounded thanks to \eqref{eq:LN}. Therefore, for any fixed $k$ there is a sequence of truncation parameters $K_1<K_2<...$ such that the tensors $A_k^{(K_1)},A_k^{(K_2)},...$ converge to some $A_k$, where $A_k$ is again left normalized. By a standard diagonal argument, one can in fact find a $k$-independent sequence of truncation parameters such that the above convergence occurs for all $k$.
Passing to the limit $K \to \infty$ in \eqref{eq:projfinite} and using the convergence of $\Psi_K$ to $\Psi$ and of  $P_{\F_k} \Psi_K$ to $P_{\F_k} \Psi$ gives eq.~\eqref{eq:proj_inf}.
Finally, letting $k \to \infty$ and using that $P_{\F_k}\Psi$ tends to $\Psi$ yields the representation \eqref{eq:prop_inf_mps}. 
 \end{proof}
 
The above proof shows that
for any  approximations $\Psi_K$ in the basis-truncated Fock space $\F_K$ which converge to $\Psi$ as $K$ tends to infinity, there exist truncation parameters $K_1<K_2<...$ such that the corresponding tensors $A^{_{(K_1)}}_k$, $A^{_{(K_2)}}_k$, ...  converge to the $A_k$ in \eqref{eq:prop_inf_mps}. See Fig. \ref{fig:matrices}.

\begin{figure}[ht!]
    \centering
    \vspace*{-1mm}
    
    \includegraphics[width = 0.195 \textwidth]{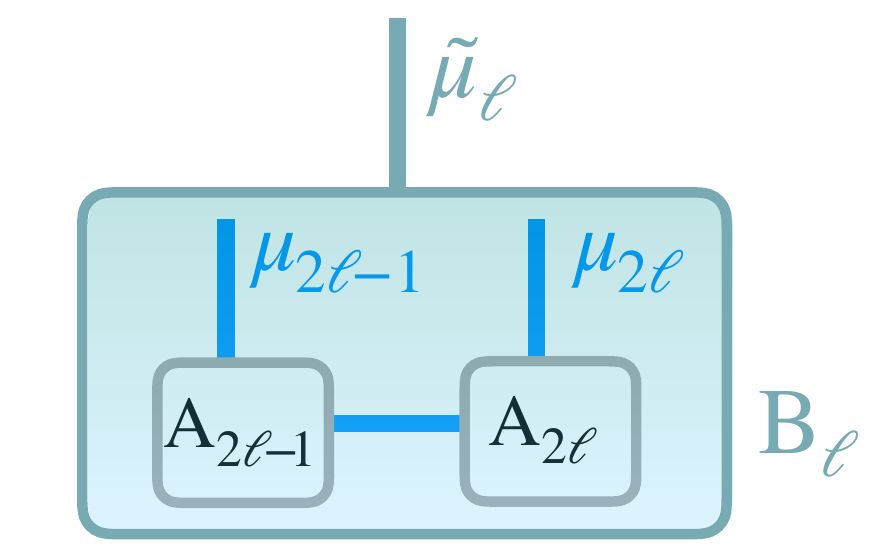}
    \\[-1mm]
    \caption{MPS matrix $B_\ell$ associated with a spatial orbital $\psi_\ell$.}
    \label{fig:pair_matrices}
\end{figure}

In quantum chemistry it is customary to absorb the spin functions into the physical variables and only use spatial orbitals as MPS nodes. The new MPS matrices subsume two old MPS matrices via
$
   B_\ell[\nu_\ell] = A_{2\ell-1}[\mu_{2\ell-1}] A_{2\ell}[\mu_{2\ell}], 
$
with $\nu_\ell$ ranging over $0,1,2,3$ corresponding to $(\mu_{2\ell-1},\mu_{2\ell})=(0,0),(1,0),(0,1),(1,1)$. See Fig. \ref{fig:pair_matrices}.
The exact representation of $\Psi$ in Theorem 
\ref{thm:mps_infinite} then takes the following form: there exist left-normalized tensors $B_k \in \C^{4^{k-1} \times 4 \times 4^{k}}$ such that, for every $k$, 
\begin{equation}\label{eq:proj_doubly_inf}
    B_1[\nu_1] \ldots  B_{k}[\nu_k]
    \mathbf{e}_k
    \widetilde{\Phi}_{\nu_1 \ldots \nu_k}
    =
    \frac{P_{\F_k} \Psi}{ \norm{P_{\F_k}\Psi}},
\end{equation}
and 
\begin{equation}\label{eq:prop_doubly_inf}
 \lim\limits_{k \to \infty} \sum_{\mathclap{\nu_1, ...,\nu_k}}  
    B_1[\nu_1] \ldots  B_{k}[\nu_k]
    \mathbf{e}_k
    \widetilde{\Phi}_{\nu_1 \ldots \nu_k}
    =
    \Psi.
\end{equation}
Here $\widetilde{\Phi}_{\nu_1,\ldots,\nu_k}$ denotes the Slater determinant in which the orbital $\psi_\ell$ is either empty or singly occupied with up or down spin or doubly occupied, according to whether $\nu_\ell=0,1,2,3$.

\section{Numerical Results} In order to validate the theory we have performed large-scale simulations on the C$_2$ molecule using the quantum chemistry density matrix renormalization group (QC-DMRG) method \cite{white_martin}. The dicarbon is modeled at 1.25 {\AA} bondlength in a frozen core cc-pVTZ \cite{dunning1989gaussian} basis. As is  customary, we then take the ensuing energy-ordered Hartree-Fock orbitals as basis functions corresponding to the MPS nodes.
Thus, one is dealing with the correlation of 
 the 8 valence electrons on 58 spatial orbitals, corresponding to a single-particle Hilbert space with dimension $d=116$ and a Fock space dimension $\approx 8 \times
 10^{34}$.
In addition, an almost twice as large frozen-core cc-pVQZ basis with 108 spatial orbitals has been investigated, yielding a single-particle Hilbert space of dimension $d=216$. We have chosen this system because of the presence of strong static and dynamic correlations; see \cite{barcza2021towards} for a recent benchmark study comparing different methods.  

We tested the convergence of the MPS matrices for the C$_2$ ground state as the basis size gets large. We focused on the singular value distribution across the bond between the matrices $A_k$ and $A_{k+1}$ for $k=8$ (corresponding to the matrices $B_k$ and $B_{k+1}$ for $k=4$). This distribution fully represents the entanglement between the Hartree-Fock occupied and virtual orbitals, up to gauge freedom.

The infinite basis limit depicted in Figure \ref{fig:matrices} can be emulated by calculating the ground state in finite bases with $K \gg k$ basis functions. To mimic passage to the large $K$ limit we present results for  various $K$ values ranging from $16$ to $d=116$. Beyond the lowest $K$ value, we also need to limit the bond dimension $D$ (note that exact representation of general states in the Fock space over 116 basis functions would require  $D\approx 3 \times 10^{17}$). Thus, we have also investigated convergence with respect to the bond dimension. 
\begin{figure}[h!]
    \centering
    \includegraphics[width = 0.5 \textwidth]{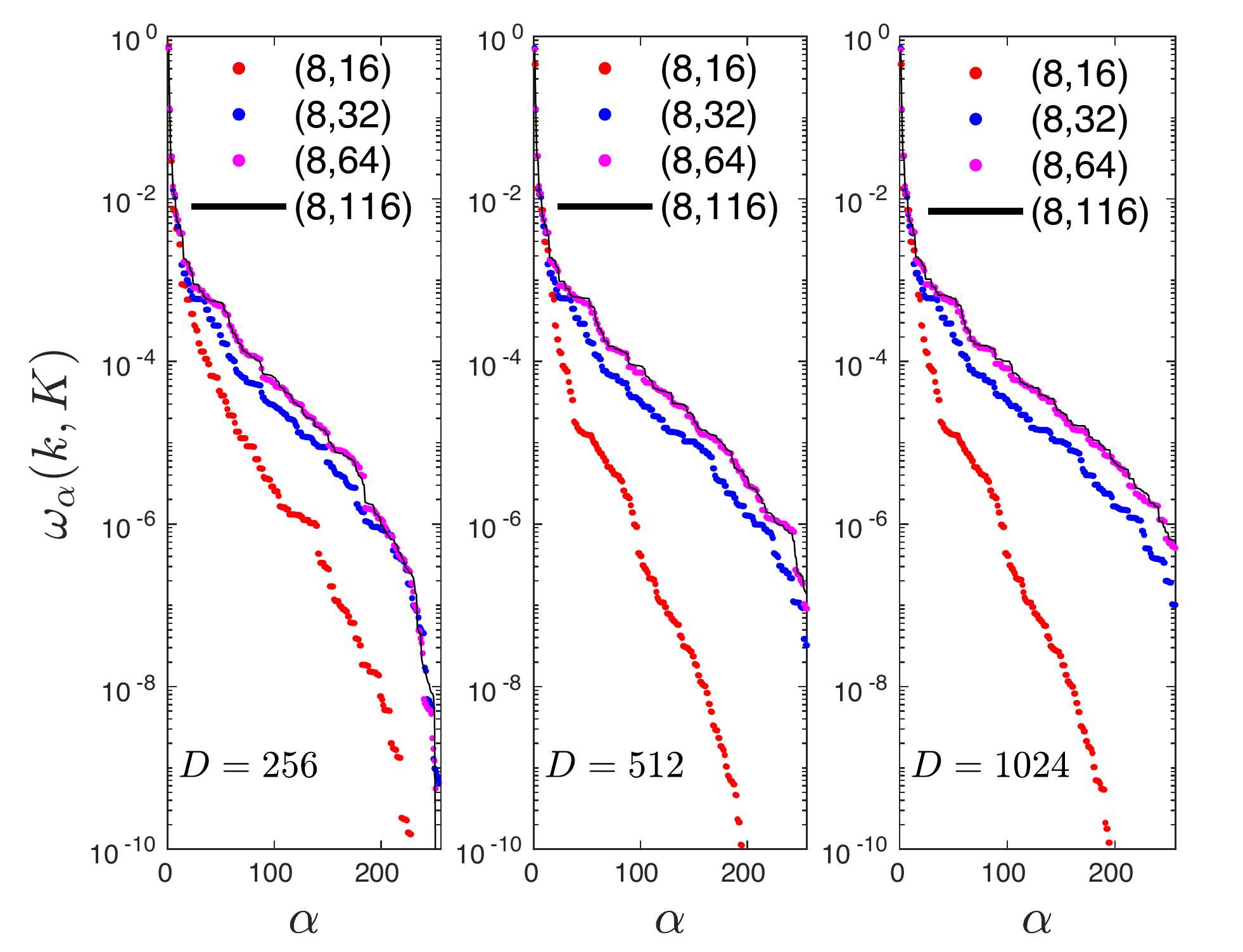}
    \vspace*{-6mm}
    
    \caption{Singular value distribution
    of the C$_2$ ground state, $\omega_\alpha(k,K)$, between core and valence space, i.e. MPS matrices $A_k$ and $A_{k+1}$ for $k=8$, and different basis sizes $K\gg k$. Inside each panel one sees the fast convergence as $K$ gets large, for fixed bond dimension $D$. Across the panels we increased the bond dimension until convergence.}
    \label{fig:numerics}
\end{figure}

The results in 
 Fig. \ref{fig:numerics}
show that the singular value distribution is already converged with respect to the size $K$ of the basis when $K\approx 64$, regardless of the bond dimension $D$.  
As for the effect of bond dimension truncation, 
we find that $D=256$ is sufficient to capture the singular values above a threshold of about $10^{-5}$, corresponding to approximately 100 singular values. The tail of the singular value distribution representing dynamic correlation effects is seen to slightly shift upwards as $D$ is increased to $512$ respectively $1024$.
Further calculations up to $D = 4096$ lead to a numerically converged spectrum, but are not shown as the difference would be hardly visible at the given axis scale. 

Results for $d=216$ up to $D=1024$ also revealed the fast convergence of the Schmidt spectrum
already at $K\approx 64$. This together with the observed exponential decay 
provides a theoretical foundation for the long-standing use of small-basis calculations in quantum chemistry.

\vspace*{1mm}

\begin{small}
{\em Acknowledgements.}
G.F. and B.R.G. have been supported by Deutsche Forschungsgemeinschaft (DFG, German Research Foundation) -- Project number 188264188/GRK1754 within the International Research Training Group IGDK 1754. 
\"OL. has been supported by the Hungarian National Research,
Development and Innovation Office (NKFIH) through Grants Nos.~K120569 and K134983, by the Quantum Information National Laboratory of Hungary, and 
by the Hans Fischer Senior Fellowship programme
funded by the Technical University of Munich -- Institute for Advanced Study.
The development of DMRG libraries has been supported by the Center for Scalable and Predictive methods for Excitation and Correlated phenomena (SPEC), funded as part of the Computational Chemical Sciences Program by the U.S.~Department of Energy (DOE), Office of Science, Office of Basic Energy Sciences, Division of Chemical Sciences, Geosciences, and Biosciences at Pacific Northwest National Laboratory.

\end{small}

\bibliographystyle{abbrv}

\end{document}